\documentclass[final,3p]{elsarticle}

\usepackage[]{natbib}
\biboptions{sort&compress}
\usepackage{amsmath}
\usepackage{amssymb}
\usepackage{url}
\usepackage{amsthm}
\usepackage{mathtools}
\usepackage{xcolor}
\usepackage{soul}

\usepackage{graphicx}
\usepackage{subcaption}
\usepackage{xcolor,colortbl}

\usepackage[colorinlistoftodos,prependcaption,textsize=tiny]{todonotes}
\usepackage[linesnumbered, ruled]{algorithm2e}

\makeatletter
	\newcommand\myparagraph{%
    \@startsection{paragraph}{4}{0mm}%
        {-\baselineskip}%
		{-0.7\baselineskip}%
        {\normalfont\normalsize\bfseries}}
\makeatother

\usepackage{float}

\begin{document}

\begin{frontmatter}
\title{Graphlets as structural fingerprints of complex networks}

\author[cas,nudz]{Anna Pidnebesna}
\ead{pidnebesna@cs.cas.cz}

\author[cas,mff]{David Hartman}
\ead{hartman@cs.cas.cz}

\author[cas,mff]{Aneta Pokorn\'{a}}
\ead{pokornaa@cs.cas.cz}

\author[cas,mff]{Daniel Trlifaj}
\ead{daniel.trlifaj@gmail.com}

\author[cas,nudz]{Jaroslav Hlinka}
\ead{hlinka@cs.cas.cz}

\address[cas]{Institute of Computer Science of the Czech Academy of Sciences, \\ Pod Vod\'{a}renskou v\v{e}\v{z}\'{i} 271/2, 182 07 Prague, Czech Republic}
\address[nudz]{National Institute of Mental Health, Topolov\'{a} 748, 250 67 Klecany, Czech Republic}
\address[mff]{Computer Science Institute of Charles University, Faculty of Mathematics and Physics, Charles University, Malostransk\'{e} n\'{a}m. 25, Prague 1, 118 00, Czech Republic}

\vspace{-1em}

  \date{\today}

\begin{abstract} 

Complex networks are often compared using selected graph-theoretical measures that capture a selected set of properties with effects ranging from local to global, such as degree, clustering or betweenness centrality. Here we introduce a structural fingerprinting framework based on graphlets: small rooted subgraphs whose distributions provide a systematic description of local-to-mesoscale topology. Across synthetic networks generated from several random graph models, graphlet fingerprints capture parameter-dependent structural differences, outperform standard graph-theoretical measures, and identify even subtle local patterns driving discrimination. We then apply the framework to empirical resting-state functional connectomes, documenting that while graphlets show superior sensitivity also to controlled topological perturbations of brain connectivity, specifically in schizophrenia–control classification they perform only comparably to classical graph-theoretical features. This is in line with the notion that schizophrenia-related alterations are dominated by spatially localized connectivity changes rather than general topological reorganization. Altogether, the generative modeling, targeted perturbations and real-world neuroimaging classification challenge position graphlets as flexible structural fingerprints of complex networks, while carefully outlining their strength and weaknesses compared to more classical graph theoretical features.

\end{abstract}


\end{frontmatter}

Complex networks are ubiquitous across domains, including social network analysis, bioinformatics, and computer science~\cite{newman2003structure,boccaletti2006complex}, and characterizing and distinguishing them is crucial for understanding variation in their structure and function~\cite{Attar2017}. A major application area is neuroscience, where network analysis of neuroimaging data has transformed the study of brain organization~\cite{bullmore2009complex,tang2023_survey}. By representing brain units-depending on the coverage and resolution, neurons, neural masses, or brain regions-as nodes and their interactions as edges, brain networks provide a formal framework for studying structural and functional connectivity~\cite{bullmore2009complex}, and for examining how distributed brain regions interact to support brain function.

A common way to characterize such networks is through graph-theoretical measures (GTMs)~\cite{bullmore2009complex, newman2010networks}. In brain networks, GTMs have helped uncover principles of brain organization, identify candidate biomarkers of neurological and psychiatric disorders, and relate network structure to behavior~\cite{stam2007graph,farahani2019application}. The increasing availability of large-scale neuroimaging datasets and mature network-analysis tools has further strengthened this approach~\cite{tang2023_survey,Rubinov2010}. In schizophrenia (SZ), for example, GTMs computed from resting-state fMRI have revealed altered topology, including changes in betweenness centrality, characteristic path length, degree, clustering coefficient, local and global efficiency, participation coefficient, and small-worldness~\cite{gao2023_graph_review,farahani2019application}. However, classical GTMs remain selective descriptors: they quantify particular selected local, global, or sometimes mesoscale properties~\cite{estrada2007characterization}, but may miss richer local-to-mesoscale configurations of network microstructure~\cite{tang2023_survey}. From a more abstract graph-theoretical perspective, the same limitation appears in the fact that even networks with a regular centrality structure may remain structurally rich, as reflected in the ongoing classification of centrality uniform graphs~\cite{gago2013betweennessuniform,hartman2024connectivity,ghanbari2023structure}.

An alternative to selecting a limited set of theoretically motivated GTMs is to evaluate a parametrized family of elementary graph patterns more systematically in the vicinity of vertices. Here we develop and evaluate such an approach based on graphlet degrees, originally introduced for biological network comparison~\cite{prvzulj2007biological}. Graphlets generalize the degree distribution of a graph to distributions of local connectivity patterns, combining aspects of adjacency-based descriptions and graph motifs~\cite{milo2002network}. The framework has been adapted for several tasks; for example, Yaveroğlu et al.~\cite{yaverouglu2014} used 11 graphlet orbits to summarize topology through the Graphlet Correlation Matrix (GCM), later applied to resting-state network classification~\cite{finotelli2021graphlet}. Related network-classification approaches have also used graphlets as unrooted motifs~\cite{janssen2012model,shervashidze2009efficient} or selected substructures such as complete subgraphs~\cite{bonato2024clique}. Beyond classification, 
graphlets are also of theoretical interest as a characteristic with the potential to fully reconstruct graphs~\cite{hartman2025reconstructing}. Their sensitivity to local and potentially mesoscale structure is particularly attractive in brain-network applications, where graph-based markers may support disease detection~\cite{amoroso2018complex,hojjati2019identification,adebisi2023brain} or treatment assessment~\cite{de2014effect}. In schizophrenia connectomics, altered homotopic connectivity and hemispheric organization have been repeatedly reported~\cite{cai2024homotopic}. Because hemispheric specialization and homotopic coordination depend on specific local circuits and cross-hemispheric pathways, graphlet-based descriptors provide a systematic way to test for asymmetries that may be averaged out by lower-dimensional summaries and thus serve as a measure of approximate symmetry~\cite{pidnebesna2025computing}.

This motivates a broader computational question: how well can network descriptors distinguish graph classes or random network generative regimes when the underlying construction rules are unknown? Many random graph models capture different aspects of network structure~\cite{newman2003structure}, and comparing empirical networks with carefully constructed models can reveal statistically significant features and suggest mechanisms shaping network organization~\cite{newman2010networks,akarca2021generative}. Conversely, when the generative rules are not known, an important task is to recognize network classes from their observed structure alone. A traditional strategy is to identify network characteristics that discriminate between generative processes and can therefore support classification of real-world networks, with GTMs often used for this purpose~\cite{Attar2017}. Graphlets provide a natural candidate for this task because they describe networks through a broad vocabulary of local topological patterns rather than through a small set of preselected scalar or nodal summaries.

In this work, we test graphlet degree distributions as structural fingerprints of complex networks. Compared with standard GTMs, graphlets provide a high-dimensional extension of node degree, with dimensionality controlled by the maximum graphlet size considered, and a systematic vocabulary of local topological patterns, allowing classifiers to identify task-relevant structures without requiring a priori selection of a small set of graph measures. We first compare graphlets and GTMs in synthetic networks generated using several random graph models with parameters that control their fine structure. We then use empirical resting-state functional connectomes to test whether graphlets detect progressive topological perturbations of brain networks and whether they provide additional information for schizophrenia-related classification. Finally, motivated by evidence for altered lateralization and homotopic connectivity in schizophrenia~\cite{cai2024homotopic}, we examine left--right hemispheric classification, SZ--HC classification within each hemisphere, and whether one hemisphere carries a stronger disease-related structural signature. 
Our contribution is therefore not the definition of graphlet degree distributions themselves, but their systematic evaluation against standard graph-theoretical measures and adjacency-based representations across nested network-comparison problems, revealing both their sensitivity to topological reorganization and their limits under edge-level mean shifts.

\section{Results}\label{results}

\subsection{Graphlet fingerprints of network structure}

We define a graphlet-based \emph{structural fingerprint} of a network as the distribution of small rooted subgraphs (graphlets) around individual vertices. For graphlets up to size $k=5$, this yields $K=73$ rooted graphlet types~\cite{prvzulj2007biological}. The resulting fingerprint records, for each node, how often each local connectivity pattern occurs in its neighborhood, i.e. its full dimension is $n \times K$ for a graph with $n$ nodes. Graphlets extend the degree distribution, which counts adjacent edges, to a multicomponent description of local-to-mesoscale topology.

\begin{figure}[h!]
\begin{center}
\includegraphics[width=1\textwidth]{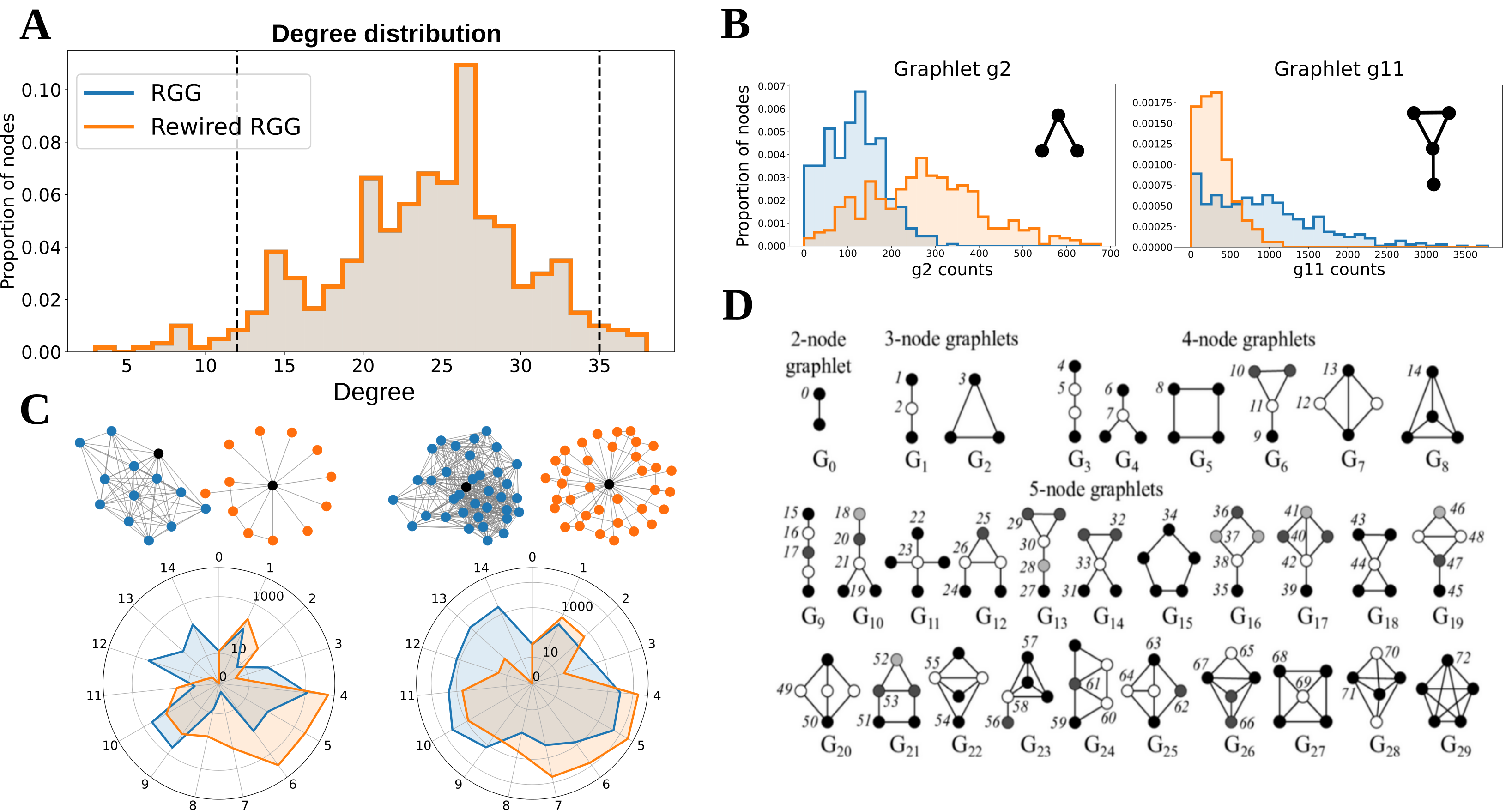}
\caption{Graphlet-based structural fingerprints. \textbf{a,} Degree distributions for a geometric random graph and its degree-preserving Maslov--Sneppen randomization~\cite{Maslov2002}. \textbf{b,} Example graphlet count distributions for two graphlets: a v-shape and a triangle with an attached node. \textbf{c,} Example structural fingerprints of two nodes with the same degree, illustrating that equal degree can correspond to different local graphlet profiles. \textbf{d,} All graphlets for $n \in {2,3,4,5}$ nodes, adapted from Pr\v{z}ulj~\cite{prvzulj2007biological}.}
\label{fig:graphletdistrib}
\end{center}
\end{figure}

Figure~\ref{fig:graphletdistrib} illustrates why such fingerprints can distinguish networks that appear similar under simpler summaries. Even vertices with the same degrees can have different graphlet profiles, and degree-preserving randomization can substantially alter the distribution of higher-order local patterns. This is consistent with studies of the small-world, where local changes can propagate into broader topological differences~\cite{watts1998collective}.

The factor $K$ can be tuned by the maximum size of the considered graphlet $k$, or by selecting subsets of graphlets when a lower-dimensional fingerprint is desirable, for example, because of dependencies among graphlet features~\cite{yaverouglu2014}. In this case, we denote the resulting reduced dimensionality by $n \times K'$.

\subsection{Graphlet fingerprints distinguish random graph models}\label{results:sim}

\begin{figure}[h!]
\begin{center}
\includegraphics[width=1\textwidth]{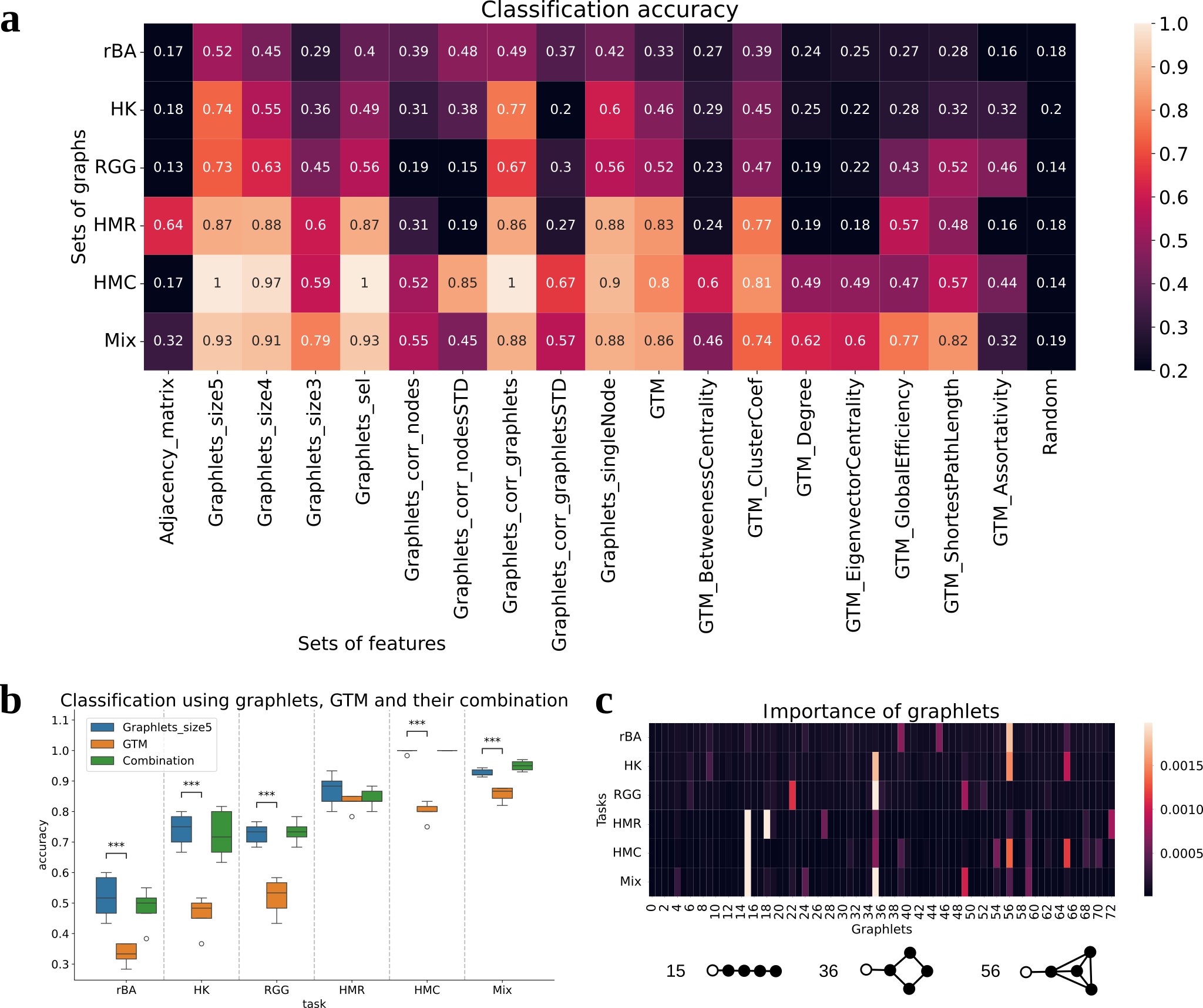}
\caption{Classification of simulated random graph models. \textbf{a,} Balanced classification accuracy for six classification tasks and 23 feature sets. Each task contains six classes defined by different model parameters; see Table~\ref{tab:graph_model_params}. \textbf{b,} Comparison of models trained on graphlet counts of size $\leq 5$, GTMs, and their combination. Graphlet and GTM models were compared using McNemar's test; *** denotes $p<0.001$. \textbf{c,} Aggregated feature importance of 73 graphlets across the six classification tasks. The most important graphlets are shown below the heatmap.}
\label{fig:results_sim}
\end{center}
\end{figure}

We first evaluated graphlet fingerprints in synthetic networks, where the generative model and its parameters are known. This provides a controlled setting in which graph classes differ in specific structural properties, such as clustering, modularity, centralization, or spatial embedding. The focus was not on identifying models that are easily detectable using standard graph metrics—such as the differences between the Erd\"{o}s–R\'{e}nyi and Barab\'{a}si-Albert models—but rather on variants of each model (defined by its parameters) that differ in more subtle structural characteristics. 
We refer to the Supplement for an example of adjacency matrix from every class. We compared graphlet-based features with standard graph-theoretical measures (GTMs), adjacency-based features, and several reduced graphlet representations; model details are provided in Section~\ref{data:sim}.

\myparagraph{Classification performance across synthetic tasks}
Figure~\ref{fig:results_sim}a summarizes balanced classification accuracies across the six synthetic tasks. The y-axis lists the classification tasks, each corresponding to six subclasses defined by different parameter values of a given graph model, and the x-axis lists the feature sets defined in Table~\ref{tab:feature_sets}. For example, the rBA task contains rewired Barab\'{a}si--Albert graphs with six levels of edge rewiring, whereas the HK task contains Holme--Kim graphs with six triangle-formation probabilities. The random-label baseline provides the expected accuracy under chance-level classification.

Graphlet fingerprints consistently performed better than, and in a few cases competitively with, the tested GTMs. The graphlet fingerprint with graphlets up to size $5$ outperformed the combined GTM feature set by approximately $2\%$--$10\%$, depending on the task. Even reduced graphlet representations retain their discriminative power -- graphlets up to size $4$ performed similarly, graphlets up to size $3$ showed only a moderate reduction, and selected graphlet families such as paths, cycles, and cliques achieved performance comparable to graphlets up to size $4$ on several tasks. Even graphlet counts from a single randomly selected node achieved high accuracy in some settings, suggesting that node-centric local information alone can capture informative signatures of global generative structure. Correlation matrices computed from graphlet distributions also characterized graph classes effectively, offering an alternative to raw graphlet count vectors.

\myparagraph{Comparison of graphlet and GTM models}
Figure~\ref{fig:results_sim}b compares GTMs, graphlet counts of size $\leq 5$, and their combination. Across the six synthetic tasks, graphlet-based classifiers achieved higher accuracy than GTM-based classifiers in five tasks. McNemar's test confirmed significant differences in all but one case ($p \leq 0.001$ for the significant comparisons). The exception was the HMR task, where GTMs and graphlets performed comparably ($p=0.11$).

Across tasks, the combined GTM+graphlet models were statistically indistinguishable from graphlet-only models ($p>0.05$). This indicates that, in these synthetic settings, the discriminative information captured by the tested GTMs was largely already represented within the graphlet fingerprint.

\myparagraph{Graphlet importance profiles}
Feature-importance analysis revealed task-specific graphlet signatures (Figure~\ref{fig:results_sim}c). Several graphlets contributed across tasks, but different graphlet types were emphasized for different generative mechanisms. Graphlet~15 (Figure~\ref{fig:graphletdistrib}d) was particularly important for the HMR, HMC, and Mix tasks. Graphlet~36 was most informative for the HK and RGG tasks, and also contributed to HMR and HMC. Graphlet~56 was most prominent for the rBA and HK tasks, with additional relevance for HMC.

These prominent graphlets are all size-five patterns in which the root is attached to a larger local structure of different density: an extended path, a cycle, or a complete graph. Their task-specific importance suggests that graphlet fingerprints distinguish generative mechanisms by identifying the local contexts in which nodes attach to sparse, cyclic, or dense neighbourhoods. Other graphlets also contributed, supporting the view that the full fingerprint captures complementary aspects of network topology.

\subsection{Graphlet fingerprints detect topological perturbations in empirical connectomes}\label{results:rewired}

\begin{figure}[h!]
\begin{center}
\includegraphics[width=1\textwidth]{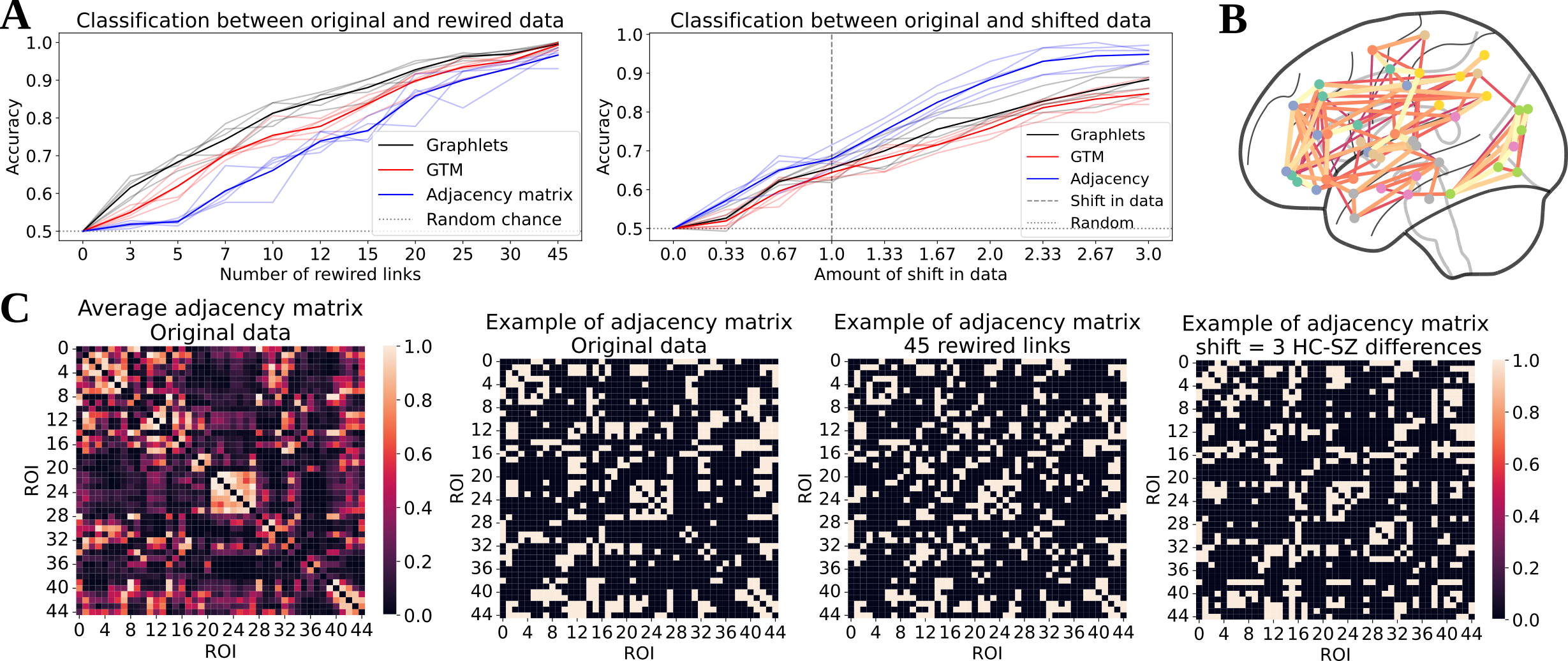}
\caption{Discrimination between empirical connectomes and their perturbed counterparts. \textbf{a,} Classification accuracy between original and rewired graphs (left) or between original and mean-shifted graphs (right) as a function of perturbation strength. Thin lines show individual cross-validation folds, and thick lines show fold-averaged trajectories. \textbf{b,} Top $10\%$ of the strongest links in the average adjacency matrix of the original data, schematically shown on the brain; darker links indicate weaker connections within this subset. \textbf{c,} Example of average adjacency matrix of the original data, original adjacency matrix, rewired adjacency matrix with 45 rewired links, and example shifted adjacency matrix (shift = 3 $\times$ SZ-HC difference).}
\label{fig:results_rewired_data}
\end{center}
\end{figure}

We next tested the detections of controlled perturbations of empirical brain networks, using two complementary perturbation types: edge rewiring and a linear shift of the group mean. For each hemisphere-level functional connectome, we generated progressively rewired versions by randomly rewiring between $0$ and $45$ edges. Separately, we computed the average adjacency matrix for HC and SZ, and shifted each individual functional connectivity by the SZ-HC difference by a coefficient ranging from 0 to 3, with real data values similar to the simulated ones for amount of shift = 1. 
Note that for this analysis, node identity is meaningful and preserved. 
At each perturbation level, Random Forest classifiers were trained to distinguish original from rewired graphs using adjacency matrices, GTMs, or graphlet counts $\leq 5$. Classification accuracy was evaluated across five cross-validation folds.

Classification accuracy increased monotonically with the perturbation strength for both perturbation types and all feature sets (Figure~\ref{fig:results_rewired_data}a), confirming that both perturbations progressively disrupted informative network structure. However, the two perturbations produced opposite sensitivity orderings. 
Under rewiring, adjacency-based classifiers showed the shallowest performance increase, GTM-based classifiers performed better, and graphlet-based classifiers achieved the highest accuracy across rewiring levels.
Under the linear mean shift, this ordering reversed: adjacency-based classifiers are the most sensitive, while graphlet- and GTM-based classifiers performed similarly to each other and required larger shifts to reach comparable accuracy.

This contrast (graphlets $>$ GTMs $>$ adjacency under rewiring; adjacency $>$ graphlets $\approx$ GTMs under mean shift) indicates that sensitivity to perturbation depends on its structural nature. Edge rewiring represents relatively local change, while preserving the rest of the graph, and is thus most readily detected by graphlet-based descriptors of local-to-mesoscale topology.
A gradual shift instead alters connectivity systematically across samples, changing it globally via a constant pattern. Such a systematic modification tends to be localized around similar nodes, which we will refer to as topographic change, across samples. This type of change is most readily detected by adjacency features that retain edge-level information, whereas graphlet and GTM summaries partially average out this signal.

\subsection{Empirical schizophrenia and hemispheric classification}\label{results:realdata}

\begin{figure}[h!]
\begin{center}
\includegraphics[width=0.8\textwidth]{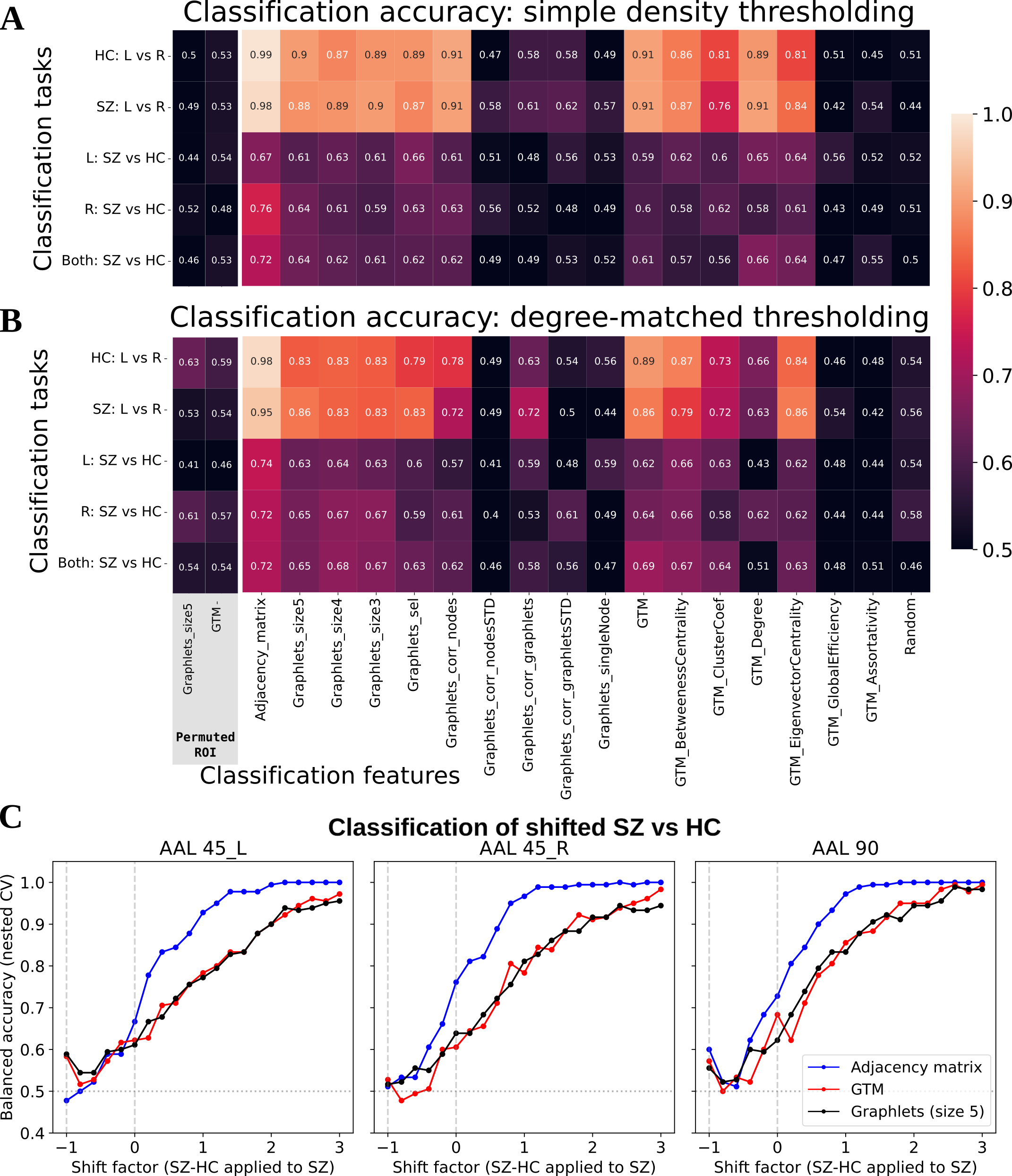}
\caption{Classification of empirical resting-state functional connectomes. Balanced classification accuracies are shown for five classification tasks and 22 feature sets. \textbf{a,} Results using proportional density thresholding. \textbf{b,} Results using degree-matched thresholding. \textbf{c,} Classification accuracy between SZ and HC after shifting SZ graphs towards or away from the HC average by a coefficient ranging from $-1$ to $3$ (shift $=0$ corresponds to unmodified data), shown separately for the left hemisphere (45 ROIs), right hemisphere (45 ROIs), and both hemispheres (90 ROIs).}
\label{fig:results_data}
\end{center}
\end{figure}

Finally, we applied the framework to empirical resting-state functional connectomes from schizophrenia patients (SZ) and healthy controls (HC) in order to classify these diagnostic groups or classify the right (R) and left (L) hemispheres. We considered two thresholding procedures: {\em proportional density thresholding} and {\em degree-matched thresholding}. The latter was included to test whether group or hemispheric differences could be explained by differences in degree distribution. These parameters resulted in a total of 10 classification tasks (5 for each thresholding) whose results are shown in Figure~\ref{fig:results_data}a,b. Across tasks, the two thresholding schemes produced similar qualitative patterns, suggesting that the main results were not driven solely by degree differences.

We also compared two ways of treating node identity. In the {\em permuted-ROI} analyses, node (ROI) labels were randomly shuffled, removing spatial correspondence between nodes while preserving topological information. In these analyses, classification accuracies were generally close to chance (see Supplement for full results). A few isolated values approached approximately $0.6$, but the overall pattern was weak and inconsistent. This indicates that unlabelled topological summaries, whether GTM- or graphlet-based, were not sufficient to reliably distinguish hemispheres or diagnostic groups in these empirical networks.

In contrast, for {\em preserved ROI} ordering, spatially specific information became highly informative. For left--right hemisphere classification, local features, including degree, clustering coefficient, betweenness centrality, and graphlet-based features, reached accuracies around $0.9$ under proportional thresholding and approximately $0.8$ under degree-matched thresholding. This pattern was similar for healthy controls and schizophrenia patients, indicating robust local hemispheric organization in both groups.

For schizophrenia-versus-control classification with preserved ROI, accuracies were more modest, typically around $0.6$--$0.65$, but consistently above chance for local characteristics including graphlets. 
With spatial organization preserved, adjacency-matrix features achieved the highest classification accuracies, outperforming both graphlet-based and GTM-based summaries. 
We compared these results to the perturbation experiment (Figure~\ref{fig:results_rewired_data}a), which contrasted two perturbation types: random rewiring, where topology was altered independently and without spatial structure for each graph, and mean-shift perturbation, where the same spatially structured change (the group-level HC–SZ difference) was applied consistently across samples. The empirical results resembled the mean-shift case rather than the rewiring case, suggesting that group difference in this dataset behaves more like a distributed shift in edge weights than like a localized reconfiguration of topology, information that adjacency features retain directly but that graph- and node-level summaries partially average away.

To test this directly, we used the mean-shift procedure introduced in Section~\ref{results:rewired}: starting from the empirical HC and SZ averages, we shifted each SZ graph towards or away from the HC average by a coefficient ranging from $-1$ to $3$, and repeated SZ-versus-HC classification at each shift level (Figure~\ref{fig:results_data}c). Shift $0$ corresponds to unmodified data; moving SZ graphs towards HC (negative shift) reduced classification accuracy to near chance, while moving them further away (positive shift) increased accuracy towards ceiling for all feature sets, with adjacency again reaching high accuracy at the smallest shift. Notably, a modest mean-only shift of SZ graphs towards HC was nearly sufficient to eliminate the classification signal, suggesting that this simple linear shift can account for most of the empirical SZ-HC difference.

Similar performance was obtained from the left hemisphere, right hemisphere, and both hemispheres however only in the preserved ROI ordering case. 
Taking into account that, on the other hand, the performance of purely global GTMs was random in both approaches and across all classifications, it further suggests that the disease-related signal was topographic but not strongly lateralized in this cohort.

Combining features did not improve performance beyond the best single one (see Supplement for details): GTM+graphlet, GTM+adjacency, and graphlet+adjacency models all performed comparably to the strongest individual representation. Thus, in the empirical classification tasks, graphlets, GTMs, and adjacency features captured partly overlapping information.

Overall, the empirical analyses show that preserving node identity and spatial correspondence is critical for uncovering discriminative structure in these functional connectomes. Within this node-preserving regime, adjacency-based features captured the most information. Graphlets provided a systematic description of local-to-mesoscale topology with sensitivity comparable to that of local graph measures. This analysis of the classification of schizophrenia showed that the differences in this dataset are more consistent with a distributed shift in edge weights than with a localized reconfiguration of the topology suggesting that effective discrimination requires descriptors beyond local-to-mesoscale topology.
Consequently, this finding reveals the nature of a potential change in network structure associated with schizophrenia.

\section{Discussion}

Across settings, graphlet fingerprints were most informative precisely when structural differences were topological rather than purely magnitude-based. In synthetic networks and under controlled rewiring of empirical connectomes, where classes differed in how edges were locally arranged, graphlets typically outperformed or at least matched standard graph-theoretical measures (GTMs) and clearly surpassed raw adjacency features. This currently makes them the superior discriminators for variants of random networks that differ only slightly in their parameters. This can be advantageous in many applications. In schizophrenia classification, however, where the discriminative signal was largely explained by a linear shift in average connectivity rather than a change in topology, this advantage disappeared: graphlets performed only comparably to local GTMs, while adjacency-based features, which retain topography directly, performed best. This result provided specific insight into the nature of network modifications in individuals with schizophrenia, which can be used in future efforts to identify a more suitable discriminator. 

The synthetic analyses established graphlets as general-purpose network descriptors that do not require a priori selection of task-relevant measures. Combining graphlets with GTMs did not improve on graphlets alone, indicating that the tested GTMs capture little information beyond what is already encoded in local connectivity patterns. Reduced graphlet representations - restricted by size or subfamily - retained much of the discriminative power, with task-specific feature-importance profiles showing that distinct generative mechanisms are fingerprinted by different local substructures. This interpretability is a practical advantage: rather than reporting a fixed panel of graph metrics, practitioners can identify which structural patterns drive discrimination in a given setting.

In schizophrenia, locally informed features - degree, clustering, graphlets, and spatially ordered adjacency - consistently outperformed purely global descriptors (global efficiency, assortativity and graphlet-based features which do not use localization information), whose performance was at chance. This is compatible with contemporary accounts of schizophrenia as a disorder of local circuit organization and synaptic gain control rather than large-scale topological  reorganization~\cite{Friston2016,Uhlhaas2010,GonzalezBurgos2012}. Separately, schizophrenia-related signal was not lateralized: accuracy was comparable across hemispheres and did not improve when both were combined, suggesting spatially distributed alterations within a broadly preserved network architecture. As anticipated above, this is consistent with our mean-shift analysis, where a simple shift of the group mean accounted for much of the empirical SZ–HC difference.

Several limitations bound these conclusions. 
We restricted graphlets to size $\leq 5$ due to computational constraint~\cite{hovcevar2014combinatorial, prvzulj2007biological}, and analyses were confined to undirected, binary, static resting-state networks. Extensions to weighted~\cite{guo2019weighted} or directed graphlets~\cite{sarajlic2016graphlet} and to task-based or dynamic connectivity are natural next steps. Finally, the schizophrenia classification accuracy, while limited in absolute terms, may reflect sample size, clinical heterogeneity, or the dominance of spatially specific effects in this cohort; importantly, however, it is comparable to state-of-the-art performance reported for schizophrenia~\cite{lei2020detecting}. Other disorders or imaging modalities may place connectomes in regimes where global topology or higher-order structure is more decisive.

Taken together, these findings position graphlet fingerprints as a flexible and interpretable extension of classical graph measures - most valuable when the discriminative structure is local-to-mesoscale and when the relevant patterns are not known in advance.

\section{Methods}
\label{data}

\subsection{Graph characteristics}
\label{graphschars}

We focus on an undirected, unweighted, and loop-free graph represented as $G=(V, E)$, where $V$ is the set of nodes (vertices) and $E$ is the set of edges (links). Unless stated otherwise, we will denote $|V| = n$. The graph is described by its adjacency matrix $A$, where each entry $a_{ij}=1$ if an edge exists between nodes $v_i$ and $v_j$ (i.e., $\{v_i,v_j\}\in E$), and $a_{ij}=0$ otherwise. The neighborhood of a node $v_i$, denoted $N_i$, is the set of nodes connected to it: $N_i=\{v_j\mid\{v_i,v_j\}\in E\}$. The degree of node $i$, denoted as $\deg_G(i)$ or simply as $k_i$, is defined as the number of its neighbors, i.e., $k_i=|N_i|$. For a subset of vertices $U\subseteq V$, we define $E(U)$ as the set of edges induced by $U$, meaning those edges with both endpoints in $U$. Given two nodes $u,v$, the path between $u$ and $v$ in a given graph $G$ is a subgraph $P_{u,v}\subseteq G$ induced by a consecutive sequence of edges, where the first edge is incident with $u$ and its other vertex is also incident with the next edge. In this way, consecutive edges always share a vertex until the last one is incident with $v$. Moreover, no vertex is repeated in the path. The length of such a path is given by the number of edges. The shortest path between $u$ and $v$ is a path of the least length. We refer to the length of the shortest path between nodes $u$ and $v$ as the {\em distance} between $u$ and $v$ and denote it as $d_{u,v}$. The longest distance between any two vertices is called {\em diameter}. More specifically, we call a graph connected if for every pair of vertices $u$ and $v$ there exists a path from $u$ to $v$.

\myparagraph{Graph-theoretical measures}
Graph-theoretical measures (GTMs) are characteristics of complex networks that capture both local (node-level) and global (graph-level) aspects of network organization. We further divide local characteristics into those based on their range within the graph: {\em characteristics with a local range}, which gather information only from the vicinity of a vertex, and {\em characteristics with a global range}, which gather information from the entire graph.  Occasionally, in accordance with~\cite{estrada2007characterization}, we refer to the range of a characteristic as {\em meso-scale} to emphasize a scope that is greater than the local scale but does not reach the global scale. The first local characteristic with a local range is represented by the node degree. For the whole graph, we calculated the {\em degree distribution} whose approximately power-law form is a defining characteristic of the famous scale-free character of complex networks~\cite{barabasi1999emergence}. It is defined as a vector of node degrees, i.e., a set of $k_i$ for all $i\in \{1,\ldots, n\}$. Generalizing the degree distribution by enabling the global range leads to  {\em eigenvector centrality}~\cite{newman2010networks}, defined using the recursive formula as
\begin{align*}
    x(v) = \frac{1}{\lambda} \sum_{t \in V} a_{v,t} x(t),
\end{align*}
where $\lambda$ is the greatest eigenvalue of the adjacency matrix $A$.

The following set of global characteristics is based on the shortest paths. The first global range characteristic, which is also a component of the small-world coefficient~\cite{watts1998collective}, is simply the average of all the shortest paths between pairs of vertices, called the {\em average shortest path}, defined as 
\begin{align*}
    L(G) = \frac{1}{n(n-1)} \sum_{u\neq v} d_{u,v}.
\end{align*}
A similar characteristic, called {\em global efficiency}, uses reciprocal values of distances, which leads to better behavior in some cases~\cite{latora2001efficient}. Global efficiency is defined as
\begin{align*}
    E(G) = \frac{1}{n(n-1)} \sum_{u\neq v} \frac{1}{d_{u,v}}
\end{align*}
Note that in situations where there is no path between vertices $u$ and $v$, and consequently $d_{u,v} = \infty$, we set $\frac{1}{d_{u,v}} = 0$. Even though it might look quite similar, behavior can be very different~\cite{latora2001efficient}. The last local characteristic with global range, based on shortest paths, is called {\em betweenness centrality}~\cite{brandes2001faster} and is defined using the equation 
\begin{align*}
    B(v) = \sum_{s\neq v\neq t}\frac{\sigma_{st}(v)}{\sigma_{st}}, 
\end{align*}
where $\sigma_{st}$ is the number of all shortest paths from $s$ to $t$, and $\sigma_{st}(v)$ is the number of all shortest paths from $s$ to $t$ that visit node $v$ at some point.

The last local characteristic that proves to be an important component of the famous small-world phenomenon is the {\em local clustering coefficient}~\cite{watts1998collective} defined as
\begin{align*}
    C(v) = \frac{2|E(N_v)|}{k_v(k_v - 1)}
\end{align*}
Note that this characteristic determines the relative number of edges (density) in the neighborhood of the vertex $v$. Similarly, we can understand it as the number of triangles with one of the nodes being $v$.

These measures were chosen to provide a comprehensive description of node importance, connectivity patterns, and overall network integration, see e.g.~\cite{farahani2019application}. 

\myparagraph{Graphlets} 
Graphlets are connected, mutually non-isomorphic subgraphs with bounded sizes (in our case of size $\leq 5$) rooted in a given node. For the sake of definition, let us consider a special type of graph called a rooted graph, which is defined as a graph with one special labeled node (called the root). Consider the set of all non-isomorphic connected rooted graphs of size max $k = 5$ with fixed ordering, see Figure~\ref{fig:graphletdistrib}d. Selected $k=5$ leads to a specific number of graphlets $K = 73$ which results in the sequence $\{H_0, H_1, \ldots, H_{K}\}$. The order of the graphlets $H_i$ is determined by the chosen order of the graphlets according to~\cite{prvzulj2007biological} and is described in Figure~\ref{fig:graphletdistrib}d. Note that choosing different $k$ leads to a different number of graphlets, e.g. $k = 4$ leads to  $K = 15$, see Figure~\ref{fig:graphletdistrib}d. The choice of graphlets up to five nodes reflects a practical and methodological compromise: graphlets of size five capture a broad range of local and mesoscale connectivity patterns while remaining computationally tractable. Although graphlets are defined as bounded rooted subgraphs, their effective topological scope depends on graph diameter and density; in small-diameter networks, graphlets of size five may capture organization extending beyond the immediate neighbourhood of a node. We therefore refer to graphlet fingerprints as local-to-mesoscale descriptors rather than purely local or global measures; the diameter distributions of the networks analysed here are reported in the Supplementary Information. Enumeration of graphlets of size six and larger is currently infeasible for graphs of the size considered here due to combinatorial complexity.

We use graphlets mainly to provide a more detailed description of the surroundings of a vertex. To this end, for the graph $G$ under study and for each vertex $v\in G$ and each graphlet $H_i$ having root $r \in H_i$, we can determine the number of subgraphs $F \in G$ such that $v\in F$ and $F$ is isomorphic to $H_i$, with the isomorphism $f:F \rightarrow H_i$ mapping $v$ to $r$, i.e. $f(v) = r$. We denote this number by the graphlet degree $\deg_{G}(H_i, v)$ or, if we have vertices of $G$ given by a numbering with the index $j$, by $g_{i,j} = \deg_{G}(H_i, j)$. Informally, this degree determines the number of distinct copies of a given graphlet $H_i$ occurring in the vicinity of a given vertex $j$. The following vector of values
\begin{align*}
    D_{*,j} = (g_{0,j}, g_{1,j}, \ldots, g_{K-1,j} )^T
\end{align*}
is called the graphlet count. If we put these counts together, we finally get a graphlet degree distribution, or graphlet distribution for short represented by a matrix $D \in \mathbb{Z}^{K,n}$.

Graphlet counts were computed as absolute frequencies without additional normalization. This design choice was intentional, as graphlets are inherently sensitive to graph density and local edge rearrangements. Normalizing graphlet counts (e.g., by degree or total subgraph count) would remove density-related information that is known to vary systematically in both generative network models and empirical brain networks. To control for trivial density effects, we therefore combined this approach with proportional and degree-matched thresholding schemes, as well as permutation-based controls.

Since the graphlet distribution matrix has dependencies along both axes (nodes and graphlets), we also consider an alternative subset of graphlets, leading to a specific subselection of rows of matrix $D$. Specifically, we considered only graphlets corresponding to paths, cycles, and cliques, referred to as \textit{Graphlets selected}. This leads to subselection of rows $\{0,1,3,4,8,14,15,34,72\}$. In another setting, instead of using the full graphlet distribution of a graph, we used the graphlet count for a single randomly selected node $j$, i.e. $D_{*,j}$, as input features called \textit{Graphlets single node}.

Lastly, we constructed correlation matrices of graphlet counts across both axes (nodes and graphlets) and used the vectorized upper triangular matrices of these as inputs to the classifier. This approach is related to the \textit{Graphlets Correlation Matrix (GCM)} introduced in~\cite{yaverouglu2014}, where the correlation was computed across nodes, and the measure was proposed to calculate the distance between the graphlet distributions. 

\subsection{Analysis overview}

\myparagraph{Task definition and simulation setup}
\label{task}

The aim of this study is to assess the ability of different graph representations (GTM, graphlets, adjacency) to classify networks according to their underlying generative mechanisms or empirical group membership. To this end, we designed a series of supervised classification tasks based on synthetic graphs, in which networks were grouped either by parameter variations within the same random graph model or by differences between distinct graph models. Detailed descriptions of the random graph models and the corresponding class definitions are provided in Section~\ref{data:sim}.

We then applied the same classification framework to empirical resting-state fMRI connectomes, focusing on discrimination between patients with schizophrenia and healthy controls, as well as between left and right hemispheres. The real dataset and associated classification tasks are described in Section~\ref{data:real}. 

To ensure comparability, all simulated networks were generated with a fixed number of $90$ nodes per graph, matching the network size of the utilized real-world dataset. For the same reason, the edge density was set to $25\%$ across all analyses. For the random graph models, only connected graphs were included to maintain consistency. For every graph model and every parameters' set, we generate $100$ graph instances.

\myparagraph{Classification features}
For each classification task, we evaluated performance using multiple feature sets. As previously described, three main categories of features were considered: graph theoretical measures (GTM), graphlet-based features, and adjacency matrix. Classification was performed separately using each individual graph theoretical measure as well as using all GTM features combined. Similarly, we conducted classification using the full set of graphlet-based features. For the adjacency matrix, we used the vectorised upper triangle. The number of features for the graph with $N$ vertices for every feature set can be found in Table~\ref{tab:feature_sets}.

\begin{table}[h]
\centering
\caption{Feature sets used for classification with their dimensionality and applicability to disconnected graphs.}
\begin{tabular}{lccc}
\hline
\textbf{Feature Set} & \textbf{Number of Features} & \textbf{Features per Graph} & \textbf{Valid for Disc.Graphs} \\
\hline
Degree distribution         & 1 value per node       & $N$                 & Yes \\
Betweenness centrality      & 1 value per node       & $N$                 & Yes \\
Eigenvector centrality      & 1 value per node       & $N$                 & Yes \\
Local clustering coefficient & 1 value per node      & $N$                 & Yes \\
Global efficiency           & 1 value per graph      & 1                   & Yes \\
Shortest path length        & $N$ values per node    & $N^2$               & No \\
All GTM together            & $N^2 + 4N + 1$ per graph & $N^2 + 4N + 1$ & Yes \\
\hline
Graphlets up to size 5            & 73 values per node     & $73N$               & Yes \\
Graphlets up to size 4            & 15 values per node     & $15N$               & Yes \\
Graphlets up to size 3            & 4 values per node      & $4N$                & Yes \\
Graphlets selected          & 8 values per node      & $8N$                & Yes \\
Graphlets single node       & 73 values per graph              & 73                  & Yes \\
\hline
Adjacency matrix & Upper triangle   & $(N-1)*(N-2)/2$  & Yes \\
\hline
\end{tabular}
\label{tab:feature_sets}
\end{table}

We conducted a multiclass classification using the Random Forest Classifier~\cite{breiman2001random} (see the technical details in Subsection~\ref{sec:meth:implementation}).
Different training and evaluation strategies were applied to artificial and real datasets. For the simulated data, the model was trained on $50$ instances and evaluated on a separate set of $50$ instances. These data contained no missing values.
For the real-world dataset, we used $K$-fold cross-validation with $K = 5$. The classifier’s performance was evaluated by computing the average accuracy across all folds.
Some graph samples in the real data were not fully connected (see~Table~1 in the Supplementary Materials), which led to missing values in shortest path length. Thus, for the real data, the shortest path lengths were excluded from the feature set.
Balanced accuracy was used as the evaluation metric for both simulated and real datasets.

\myparagraph{Comparison of Graphlet- and GTM-Based Classification Performance}

To evaluate whether graphlet-based features provide superior discriminative power compared to standard graph-theoretical measures (GTM), we performed a systematic classification comparison across all six experimental tasks. Each task represented a distinct classification scenario, and for each, models were trained separately using three feature sets: (\textit{i}) GTM only, (\textit{ii}) graphlet-based features (size $\leq5$), and (\textit{iii}) the combined feature set (GTM + graphlets).

For the simulation study, data were randomly split into independent training and testing subsets (50/50 ratio). To estimate variability in model performance, the held-out test data were further partitioned into five equal, non-overlapping subsets, yielding five independent accuracy estimates per model and per task. Mean accuracies and variability across these five repetitions were visualised using boxplots.

For statistical comparison of classifier performance, we applied McNemar’s test to paired classification outcomes obtained from GTM and graphlet models, 
thereby assessing whether the two feature representations differed significantly in their ability to correctly classify individual samples \cite{McNemar1947, dietterich1998approximate}. McNemar’s test assesses whether two classifiers differ significantly in their proportions of correctly and incorrectly classified instances by focusing on samples where their predictions disagree.
The same procedure was used to compare the combined (GTM + graphlet) model against the graphlet-only model, allowing us to test whether adding GTM features provided any incremental benefit beyond the information contained in graphlets.

\myparagraph{Feature Importance Analysis for Graphlet-Based Models}

To identify which graphlet patterns contributed most strongly to classification performance, we computed feature importances from the fitted classifier for each classification task. Importances were estimated using permutation-based importance scores \cite{breiman2001random}, which quantify the mean decrease in model accuracy upon random shuffling of each feature while holding all others fixed. This approach provides an interpretable measure of the unique contribution of each feature to predictive performance.

Permutation importance was computed on the held-out test data, with ten random permutations per feature to ensure stability. The resulting importance vectors were normalized (z-scored) within each task and aggregated by graphlet type and size (graphlets up to five nodes). To visualize task-specific feature relevance, we organized graphlet features along one axis and classification tasks along the other, producing a two-dimensional heatmap where color intensity reflected relative feature importance. This allowed direct comparison of motif-level contributions across tasks and facilitated interpretation of whether distinct classes of local topology (e.g., triangles vs. four-node stars) were preferentially informative for particular classification contrasts.

\myparagraph{Implementations notes}
\label{sec:meth:implementation}

All computational analyses were performed using Python version 3.12.1. Most of the graph models and graph-theoretical metrics were computed using the \texttt{networkx} library (version 3.4.2); for HMR and HMC models, the implementation was adopted from~\cite {pyGAlib}. Classification tasks were carried out using the \texttt{scikit-learn} library (version 1.3.0). 
We conducted a multiclass classification using the Random Forest Classifier \texttt{sklearn.ensemble.RandomForestClassifier}. To ensure replicability, the random seed was set to $42$. The classifier was trained using the default parameters. 
For classification, per-node feature matrices (e.g., the $n \times K$ graphlet distribution, or per-node GTMs) were flattened into a single feature vector per graph by concatenating values across nodes in a fixed, spatially consistent node ordering (for empirical data, ROI order; for synthetic data, node index at generation). This preserves node identity across samples but yields high-dimensional feature vectors relative to sample size. 
We rely on Random Forest's built-in feature subsampling and bagging to manage this $p \gg n$ regime; no additional dimensionality reduction was applied except in the explicitly reduced feature sets (Graphlets selected, Graphlets single node, Graphlets correlation) described above.
Graphlet computations were performed using the ORCA C++ library~\cite{hovcevar2014combinatorial}, which was interfaced with Python via custom wrapper code.

\subsection{Simulated data}
\label{data:sim}
We start the description from the Barabási-Albert (BA) model~\cite{barabasi1999emergence}. Although the original BA model is not used directly in our experiments, we employ two modified versions that build on its structure. The BA model is a widely used random graph model designed to capture the scale-free degree distribution frequently observed in real-world complex networks. This property affects many characteristics, including the degree distribution, eigenvalue centrality, and the more general graphlet distribution. 

The generative process begins with a small, connected seed graph, which is incrementally expanded by adding new nodes iteratively. Each newly added node is connected to a fixed number (parameter of the model) of existing nodes, with connection probabilities proportional to target node degrees — a mechanism known as preferential attachment. This results in networks where a few nodes (hubs) accumulate disproportionately many connections, reflecting real-world connectivity patterns. Since many networks have been shown to be modeled well to some extent by the BA model or its variants, this model is a good starting point for testing the approach on real networks.

\myparagraph{Rewired Barabási-Albert Model}
The first variant used in this study is the rewired BA model, which introduces stochastic perturbations to the original structure through edge rewiring, a process commonly used to assess the robustness of network characteristics to imperfect or perturbed network structure~\cite{borgatti2006robustness,pidnebesna2025computing,Maslov2002}. Specifically, a fraction of edges in a BA graph is selected uniformly at random and rewired by removing an existing edge and reconnecting a pair of previously unconnected nodes, chosen uniformly at random. 
We define six graph classes based on the proportion of rewired edges, with the rewiring levels regularly sampled between $10\%$ and $45\%$, resulting in the following values: $10\%$, $17\%$, $24\%$, $31\%$, $38\%$, and $45\%$. 

\myparagraph{Holme-Kim Model}
The second Barabási-Albert model variant utilized in this study is the Holme-Kim (HK) model~\cite{Holme2002}, which extends the BA model by incorporating a mechanism to enhance local clustering, thereby increasing the prevalence of triangles in the network. The HK model retains the preferential attachment principle of the BA model but introduces a triangle-closing step: after attaching a new node to an existing one via preferential attachment, with a specified probability (parameter of the model), it also connects to one of the neighbors of the newly attached node, forming a triangle. This process results in networks that more accurately reflect the clustering characteristics observed in many real-world systems. 
To examine the influence of clustering on classification performance, we generated six classes of graphs by varying the triangle formation probability. The selected probabilities were $0.15$, $0.30$, $0.45$, $0.60$, $0.75$, and $0.90$. 

\myparagraph{Random Geometric Graph Model}

The Random Geometric Graph (RGG) model generates networks by placing nodes uniformly at random within a geometric space and connecting pairs of nodes whose Euclidean distance falls below a fixed threshold~\cite{penrose2003random}. This model is commonly used to represent spatially embedded systems, where connectivity depends on physical proximity rather than abstract probabilities. The dimensionality of the embedding space plays a critical role in shaping the network structure: higher dimensions generally lead to more homogeneous connectivity and reduced clustering, while lower dimensions result in more locally structured graphs.
In this study, we defined six graph classes based on the dimensionality of the unit hypercube in which the nodes were embedded. Specifically, we used dimensions $3$, $5$, $7$, $9$, $11$, and $13$, with the corresponding connection threshold subsequently applied to maintain a constant edge density.

\myparagraph{Hierarchical Modular Random Model}

The Hierarchical Modular Random (HMR) model generates networks with modular, nested community structures, reflecting hierarchical organization commonly observed in biological and social systems~\cite{ZamoraLopez2016}. In this model, nodes are arranged into a hierarchy of modules, with connection probabilities between nodes determined by their position in the hierarchy. This framework allows for the creation of graphs that exhibit both local clustering within modules and long-range connections across the hierarchy, resulting in rich and realistic topological features.

For our experiments, we employed an HMR variant defined by a fixed modular structure comprising three levels: 2 top-level modules, each containing 3 mid-level submodules, which in turn consist of 15 nodes each—resulting in graphs with 90 nodes in total. Six graph classes were generated by varying the intra- and inter-module connection probabilities, denoted by parameters $(a, b, c)$, which respectively control edge densities within lowest-level submodules, between submodules of the same mid-level module, and between different top-level modules. The values of parameters used were: $(3,10.2,9)$, $(3, 7.2, 12)$, $(5, 8.2, 9)$, $(5,5.2,12)$, $(7,6.2, 9)$, and $(7,3.2,12)$ for classes 1 to 6 respectively.

\myparagraph{Hierarchical Modular Centralised Model}

The Hierarchical Modular Centralised (HMC) model generates networks that combine hierarchical modular organization with centralized connectivity, reflecting empirical observations in brain networks where local specialization coexists with global integration via hub nodes~\cite{ZamoraLopez2016}. The model extends the HMR framework by introducing preferential attachment mechanisms that concentrate inter-module connections onto a subset of central nodes, yielding hub-dominated long-range connectivity.

Networks were organized into three modules of 30 nodes each, defining a fixed hierarchical structure. Average within-submodule connectivity was 16 links per node, while inter-module connectivity averaged 6.3 links per node. Centralization was controlled by two preferential attachment exponents governing inter- and intra-module connectivity, respectively. Six parameter combinations were examined,
$(1.8,2)$, $(2.2,2)$, $(4,2)$, $(1.8,25)$, $(2.2,25)$, and $(4,25)$, spanning a range from distributed to strongly centralized, hub-dominated topologies.

\myparagraph{Mixed model task}
Finally, we constructed a \emph{mixed} classification task that combined graph instances drawn from multiple generative mechanisms. This task was designed to evaluate the ability of different feature sets to discriminate between qualitatively distinct network architectures rather than parametric variants of a single model. Specifically, the task included Erd\H{o}s--R\'enyi (ER) random graphs with $N=90$ nodes and edge probability chosen to yield an expected edge density of $25\%$, together with one representative class from each previously described random graph family. These included the rewired Barab\'asi--Albert model with rewiring probability $0.24$, the Holme--Kim model with triangle formation probability $0.45$, a random geometric graph embedded in a $7$-dimensional unit cube, the hierarchical modular random (HMR) model with intra- and inter-module connection probabilities $(5, 8.2, 9)$, and the hierarchical modular configuration (HMC) model with intra- and inter-module degree exponents $(4, 2)$. By combining structurally diverse graph models within a single task, this setting probes the robustness of graph descriptors to heterogeneous generative processes.

\begin{table}[h]
\centering
\caption{Summary of graph generation parameters for each random graph model. Each class contains 100 graph instances. All graphs have 90 nodes and 25\% edge density.}
\label{tab:graph_model_params}
\begin{tabular}{l l l}
\hline
\textbf{Model} & \textbf{Parameter(s)} & \textbf{Class Values} \\
\hline
rBA & Percentage of rewired edges & 0.10, 0.17, 0.24, 0.31, 0.38, 0.45 \\
HK & Triangle formation probability & 0.15, 0.30, 0.45, 0.60, 0.75, 0.90 \\
RGG & Dimensionality of unit cube & 3, 5, 7, 9, 11, 13 \\
HMR & Intra/inter-module probabilities $(a,b,c)$ & $(3,10.2,9)$, $(3, 7.2, 12)$, $(5, 8.2, 9)$,  \\
 & 2 x 3 x 15 modules & $(5,5.2,12)$, $(7,6.2, 9)$, $(7,3.2,12)$ \\
HMC & Intra/inter-module exponents & $(1.8,2)$, $(2.2,2)$, $(4,2)$,  $(1.8,25)$, $(2.2,25)$, $(4,25)$ \\
 Mix & ER + Class 3 from presented graph models & ER, rBA, HK, RGG, HMR, HMC \\
\hline
\end{tabular}
\end{table}

\subsection{Real data}
\label{data:real}

\myparagraph{Data collection and preprocessing}
The human brain is a dynamic system whose functional organization can be explored using functional magnetic resonance imaging (fMRI), a non-invasive technique that measures blood-oxygen-level-dependent (BOLD) signal fluctuations. Resting-state fMRI (rs-fMRI), in particular, allows for the investigation of spontaneous brain activity in the absence of specific tasks, revealing patterns of functional connectivity (FC) that reflect the brain's intrinsic organization.

In this study, we analyzed resting-state fMRI data obtained from 90 schizophrenia (SZ) patients and 90 healthy controls (HC). All data were acquired using a 3T Siemens Magnetom Trio scanner at the Institute for Clinical and Experimental Medicine (IKEM), Prague. Preprocessing of the fMRI data followed standardized pipelines, which included motion correction, spatial normalization, and temporal filtering; full methodological details are provided in~\cite{Tomecek2024}.

\myparagraph{Graphs from functional connectivity}
To construct functional brain networks, we extracted regional time-series from 90 regions of interest (ROIs) defined by the Automated Anatomical Labeling (AAL) atlas~\cite{Tzourio-Mazoyer2002}. Functional connectivity matrices were computed for each subject by calculating pairwise Pearson correlation coefficients between all ROI time-series as this correlation measure was found to be sufficient for network construction~\cite{hartman2011role}. 
The resulting correlation matrices were then thresholded to retain only the $25\%$ strongest connections after removing the diagonal, ensuring a consistent edge density across subjects ({\em proportional thresholding}). This procedure yielded binary, undirected graphs representing individual functional connectomes.

In addition to proportional thresholding, we implemented a {\em degree-matched thresholding} procedure to further control for topological biases across groups. Specifically, for each subject’s FC matrix, we constructed a graph by iteratively adding the strongest edges while constraining node degrees to follow a target degree distribution. The target distribution was obtained by averaging the degree sequences across all graphs derived with the proportional thresholding method. Edges were added in descending order of correlation value until the desired global density was reached, while ensuring that the degree of each node did not exceed its target value. This approach yields graphs with approximately preserved edge density and group-consistent degree distributions, thereby reducing variability due to thresholding and enabling more balanced comparisons of network features across subjects.

To determine whether the observed class differences arise from local (spatially specific) or global connectivity patterns, we performed two analyses: one using spatially {\em ordered ROIs}, in which we preserve particular IDs of nodes, and another using randomly {\em permuted ROI} assignments, where we remove node identity.

\myparagraph{Instances for classification tasks}
We performed several classification tasks using the described data. In particular, we worked with the following classes: left versus right hemisphere for control group (HC: L vs R, 45 ROI), left versus right hemisphere for patients group (SZ: L vs R, 45 ROI), controls versus patients using left and right hemisphere separately (L: HC vs SZ, 45 ROIs; and R: HC vs SZ, 45 ROIs), controls versus patients using both hemispheres (Both: HC vs SZ, 90 ROI).
The information about the number of subjects, the number of regions, and the number of connected graphs for the described data can be found in the {Supplementary Material, Table~1}. 

\myparagraph{Rewired brain data model}
In addition, to assess the sensitivity of the extracted network features to perturbations in the empirical data, we generated rewired functional connectivity (FC) matrices by gradually permuting edges. Specifically, we considered hemisphere-level networks (45 ROIs) for both control and patient groups, resulting in a total of 360 matrices. For each matrix, we progressively rewired between 0 and 45 randomly selected edges, with the rewiring performed independently for each subject’s network. This procedure allowed us to create controlled perturbations of the real connectivity structure while preserving the overall network density. The rewired datasets were then used in classification tasks to evaluate the ability of the features to discriminate between original and perturbed networks.

\myparagraph{Mean-shifted brain data model}
As a complementary perturbation, we tested sensitivity to a linear shift of the group mean rather than a redistribution of edges. For each hemisphere-level network (45 ROIs), we computed the average adjacency matrix separately for the HC and SZ groups and defined their difference as the shift direction. Each individual graph was then shifted by adding this difference matrix, scaled by a coefficient c, to its original adjacency matrix, followed by re-thresholding to preserve the original edge density. We first evaluated classification between original (c=0) and shifted graphs at coefficients ranging from 0 to 3, applied uniformly across both groups, to test the general sensitivity of each feature set to this type of perturbation (Figure~\ref{fig:results_rewired_data}a). We then applied the same procedure specifically to SZ graphs, shifting them towards or away from the HC average using coefficients ranging from -1 to 3, and repeated SZ-versus-HC classification at each shift level (Figure~\ref{fig:results_data}c). Because the shift direction was itself estimated from the group averages, this analysis was performed using nested cross-validation, with the HC-SZ difference vector re-estimated on each training fold and applied only to the corresponding held-out test data, avoiding circularity between the estimated shift direction and its evaluation.

\section*{Data availability}

The simulated random graph datasets generated in this study, together with the derived feature matrices and source data underlying the main figures, will be deposited in Zenodo upon publication of the journal article. The derived, anonymized functional connectivity matrices used for the empirical connectome analyses will also be deposited in Zenodo upon publication, subject to institutional and ethical approval for sharing derived neuroimaging data. The raw resting-state fMRI data and individual-level clinical or demographic information are not publicly available because they contain sensitive human-participant data and are subject to institutional, ethical and data-protection restrictions. Access to restricted data may be considered by the corresponding author upon reasonable request and subject to approval by the relevant data controller and ethics committee.

\section*{Code availability}

The custom code used to generate the synthetic networks, compute graphlet- and graph-theoretical features, perform the classification analyses, run the perturbation experiments and reproduce the main figures will be deposited in a public GitHub repository and archived in Zenodo upon journal article publication. The repository will include documentation, analysis scripts and information on the software environment required to reproduce the analyses. Graphlet computations used the ORCA C++ implementation interfaced with Python via custom wrapper code; other analyses used standard Python packages as described in the Methods.

\section*{Funding}

The authors were supported by the Czech Science Foundation Grant No. 23-07074S and by the ERDF Project Brain dynamics, No. CZ.02.01.01/00/22\_008/0004643.

\section*{Author contributions}

Conceptualization: A.Pi., D.H. and J.H. Methodology: all authors. Software and implementation: A.Pi., D.H., A.Po. and D.T. Simulations and formal analysis: A.Pi., J.H., D.H. and D.T., Empirical data: J.H. Statistical design, control analyses and model comparisons: A.Pi., D.H., and J.H.; Visualization: A.Pi. Interpretation: all authors. Writing---original draft: A.Pi., D.H., J.H. Writing---review and editing: all authors. Supervision: D.H. and J.H. Funding acquisition: J.H. annd D.H. All authors approved the final version.

\section*{Competing interests}

The authors declare no competing interests.

\section*{Ethics approval}

The empirical resting-state fMRI data analysed in this study were collected in previous studies approved by the relevant institutional ethics committee(s). All participants provided written informed consent in accordance with the approved study protocol and applicable regulations. The present analyses used anonymized data and were conducted in accordance with the institutional approvals governing secondary analysis of these data.

\section*{Generative AI statement}

During manuscript preparation, the authors used generative AI tools, including ChatGPT (OpenAI) and Claude (Anthropic), to assist with language editing, condensation, clarity, organization and formulation of explanatory text. These tools were not used to generate data, perform statistical analyses, produce the reported numerical results or make autonomous scientific conclusions. All AI-assisted text was critically reviewed, edited and approved by the authors, who take full responsibility for the final content of the manuscript.

\end{document}